\documentclass[
 reprint,
superscriptaddress,
 amsmath,
 amssymb,
 aps,
 prb,
]{revtex4-2}

\usepackage{graphicx}
\usepackage{dcolumn}
\usepackage{bm}
\usepackage[hidelinks]{hyperref}

\hypersetup{
  colorlinks   = true,
  urlcolor   = blue,
  citecolor   = blue
}

\begin{document}

\preprint{APS/123-QED}

\title{Raman signature of cation vacancies in rare-earth nitrides}

\author{M.~Markwitz}\email{martin.markwitz@vuw.ac.nz}
 \affiliation{Robinson Research Institute, Victoria University of Wellington, P.O. Box 33436, Petone 5046, New Zealand}
 \affiliation{The MacDiarmid Institute for Advanced Materials and Nanotechnology, Victoria University of Wellington, P.O. Box 600, Wellington 6140, New Zealand}
\author{K.~Van~Koughnet}
 \affiliation{The MacDiarmid Institute for Advanced Materials and Nanotechnology, Victoria University of Wellington, P.O. Box 600, Wellington 6140, New Zealand}
 \affiliation{School of Chemical and Physical Sciences, Victoria University of Wellington, P.O. Box 600, Wellington 6140, New Zealand}
 \author{K.~Kneisel}
 \affiliation{The MacDiarmid Institute for Advanced Materials and Nanotechnology, Victoria University of Wellington, P.O. Box 600, Wellington 6140, New Zealand}
 \affiliation{School of Chemical and Physical Sciences, Victoria University of Wellington, P.O. Box 600, Wellington 6140, New Zealand}
\author{W.~F.~Holmes-Hewett}
 \affiliation{Robinson Research Institute, Victoria University of Wellington, P.O. Box 33436, Petone 5046, New Zealand}
 \affiliation{The MacDiarmid Institute for Advanced Materials and Nanotechnology, Victoria University of Wellington, P.O. Box 600, Wellington 6140, New Zealand}
 \author{F.~Natali}
 \affiliation{The MacDiarmid Institute for Advanced Materials and Nanotechnology, Victoria University of Wellington, P.O. Box 600, Wellington 6140, New Zealand}
 \affiliation{School of Chemical and Physical Sciences, Victoria University of Wellington, P.O. Box 600, Wellington 6140, New Zealand}
 \affiliation{Liquium Ltd., Laby Building, Victoria University of Wellington, Wellington 6012, New Zealand}
\author{E. X. M. Trewick}
 \affiliation{The MacDiarmid Institute for Advanced Materials and Nanotechnology, Victoria University of Wellington, P.O. Box 600, Wellington 6140, New Zealand}
 \affiliation{School of Chemical and Physical Sciences, Victoria University of Wellington, P.O. Box 600, Wellington 6140, New Zealand}
\author{L.~Porteous}
 \affiliation{School of Chemical and Physical Sciences, Victoria University of Wellington, P.O. Box 600, Wellington 6140, New Zealand}
\author{B.~J.~Ruck}
 \affiliation{The MacDiarmid Institute for Advanced Materials and Nanotechnology, Victoria University of Wellington, P.O. Box 600, Wellington 6140, New Zealand}
 \affiliation{School of Chemical and Physical Sciences, Victoria University of Wellington, P.O. Box 600, Wellington 6140, New Zealand}
\author{H.~J.~Trodahl}
 \affiliation{School of Chemical and Physical Sciences, Victoria University of Wellington, P.O. Box 600, Wellington 6140, New Zealand}
 
\date{\today}

\begin{abstract}We report a coordinated Raman/computation study of the rare-earth nitrides, a series of intrinsic ferromagnetic semiconductors, to reveal the presence of cation vacancies. Their presence is signaled by a Raman-active vibrational mode at 1100-1400~cm$^{-1}$, rising steadily as the lattice contracts across the series. The mode's frequency is in excellent agreement with the computed breathing-mode vibration of the six nitrogen ions surrounding cation vacancies. The discovery of such cation vacancies opens the door for hole doping that has so far been lacking in the exploitation of rare-earth nitrides. 

\end{abstract}

\maketitle

\section{Introduction}

The rare-earth mononitrides (\textit{Ln}N, \textit{Ln} a lanthanide) form a mutually epitaxy-compatible series of 15 intrinsic ferromagnetic semiconductors with promise for mixed superconductor-spintronics~\cite{Natali2013,Senapati2011,Pal2013,Caruso2019,Ahmed2020,Pot2023}. The combined spin and orbital magnetic moments within their 4\textit{f} shell precipitate enormously varied magnetic states and properties that have been gradually investigated, but the control over their electronic properties remains a largely unsolved problem. In particular a small formation energy for nitrogen vacancies leave thin films heavily electron doped and there is as of yet no well-characterized compensating acceptor without the incorporation of Mg, the use of which is undesirable in vacuum systems. Here we report Raman measurements that surprisingly suggest the way to the first hole-doped \textit{Ln}N by utilizing intrinsic cation vacancy defects.

The \textit{Ln}N adopt the NaCl structure with lattice constants varying from 0.530~nm in LaN to 0.476~nm in LuN~\cite{Natali2013}. Most current work focuses on films grown in a nitrogen atmosphere, relying on catalytic breaking of the N$_2$ triple bond at a clean \textit{Ln} surface~\cite{Ullstad2019,chan2020facile,chan2023}, though \textit{Ln}N powders formed in a nitrogen atmosphere are also reported~\cite{Kneisel2024}. There have been only very few reports of near-macroscopic single crystals~\cite{wachter2012physical, AlAtabi2018, Degiorgi1990}, such that the vast majority of studies have been performed on polycrystalline powders or thin films. Investigating departures from stoichiometric crystalline order for the exploration of engineered defects for carrier doping remains a challenge. Within this space Raman spectroscopy is supremely efficient, even encouraging its use in situations such as in rock-salt crystals adopted by the \textit{Ln}N, where conventional first-order Raman activity is not allowed. 

\begin{figure}
    \centering
    \includegraphics[width=\linewidth]{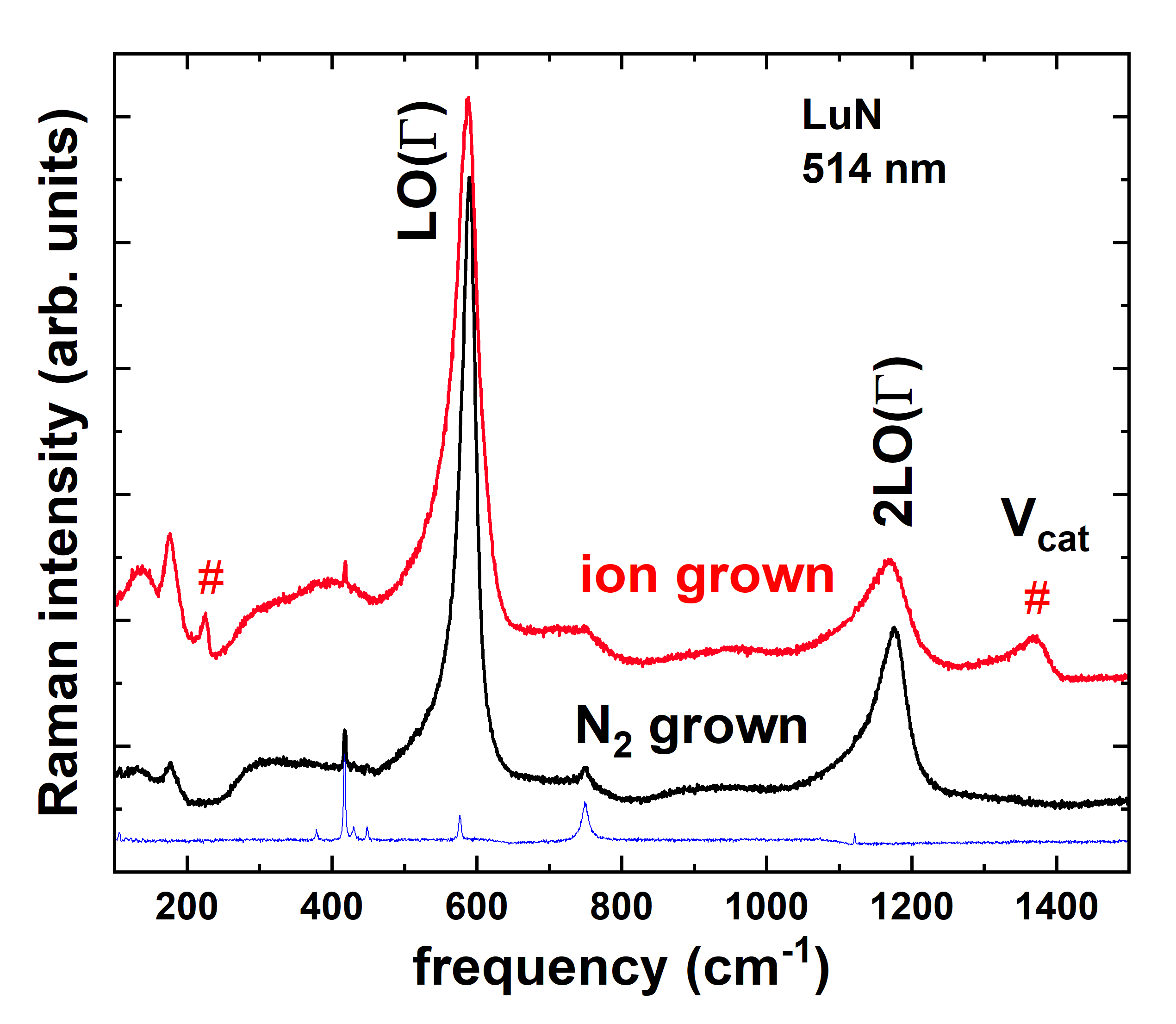}
    \caption{Raman spectra of LuN films grown with nitrogen ions (red) and molecules (black). The most prominent feature is, as labeled, a Fr\"ohlich interaction-activated LO($\Gamma$) signal, seen also at double the frequency and labeled as 2LO($\Gamma$). The low-level broad scattering below 750~cm$^{-1}$ in both spectra is a mix of intrinsic two-phonon and extrinsic defect-activated scattering. The features marked with hashes are found only when the \textit{Ln}N are grown in activated nitrogen, signaling specific defects that were uniquely introduced by ions. It is the feature at $\sim$1380 cm$^{-1}$ that is identified here as a breathing mode centered on a cation vacancy and thus labeled appropriately as V$_{\text{cat}}$. The blue line is the low-intensity spectrum of the bare sapphire substrate.}
    \label{fig:LuNcompare}
\end{figure}

There are several examples of Raman spectroscopy as a characterization tool for the \textit{Ln}N~\cite{Granville2009,Upadhya2022,VanKoughnet2023}, recently yielding separate identification of intrinsic and defect-activated Raman spectra of two of the series, GdN and LuN~\cite{VanKoughnet2023}. The Raman signal of LuN films seen in Fig.~\ref{fig:LuNcompare} show a strong intrinsic LO($\Gamma$) feature, a longitudinal optical phonon mode at the $\Gamma$ point, at $\sim$540-590~cm$^{-1}$ and its double, 2LO($\Gamma$), at $\sim$1080-1180~cm$^{-1}$, both activated by an unconventional exciton-phonon Fr\"ohlich interaction~\cite{merlin1978multiphonon}. Van Koughnet \textit{et al.}~\cite{VanKoughnet2023} discuss in more detail the identification and origin of the Raman signal of the \textit{Ln}N series to be attributed to the Fr\"ohlich-excited LO($\Gamma$) mode, rather than a defect-activated LO(L) vibrational mode suggested in an early study~\cite{Granville2009}. The other weaker signals below 750~cm$^{-1}$ arise from a combination of intrinsic two-phonon and extrinsic point defect-activated scattering~\cite{VanKoughnet2023}. Raman studies of the other 14 \textit{Ln}N fit within this picture, though with the usual differences among the relative strengths resulting from the mix of crystal defects~\cite{Upadhya2022,Granville2009,VanKoughnet2023}. The films used in Fig.~\ref{fig:LuNcompare} were grown in either ionized or molecular nitrogen, showing the respective presence or absence of defect-activated Raman signal modes in the \textit{Ln}N. These peaks are signed with hashes, the high-frequency mode which lays $\sim$200~cm$^{-1}$ higher than the 2LO($\Gamma$). A similar feature has been noted in \textit{Ln}N powders formed by ball milling in N$_2$~\cite{Kneisel2024}; evidently both ball milling and ion-grown films harbor the defect.

This paper focuses on the identification of the defect responsible for the feature. Its high frequency immediately suggests a mode involving the vibration of nitrogen, for which an obvious candidate is a breathing-mode of the six N ions surrounding a cation vacancy. It is common that such breathing mode vibrations modulate the defect center's polarizability to provide a strong Raman cross section, motivating the computational study of that breathing-mode frequency which we report here. In anticipation of the evidence as follows that the feature is the cation-vacancy breathing mode, we label it as V$_{\text{cat}}$. As mentioned, one of the vexing problems that limits electronic developments for the \textit{Ln}N is the lack of a hole dopant. The activation energy of nitrogen \textit{anion} vacancies (V$_{\text{an}}$), each releasing up to three electrons, is so small as to ensure electron doping in any material grown at the elevated temperatures required for epitaxial growth~\cite{Punya2011,porat2024tuneable}. Mg substitution to $\leq$1\% is known to compensate the V$_{\text{an}}$, but the high vapor pressure of Mg discourages its introduction into any molecular beam epitaxy (MBE) vacuum chamber~\cite{Lee2015,Lee2018}. One might hope that \textit{cation} vacancies are capable of offering a compensating hole dopant.

The following section (II) describes the experimental Raman study of the \textit{Ln}N, starting with the preparation and characterization of thin film and powder samples. We then describe the Raman measurements and follow with the results. The important results concern the presence or absence of the the V$_{\text{cat}}$ breathing mode in the \textit{Ln}N of differing atomic number (Z) across the \textit{Ln} series. In Section III we report density functional theory (DFT) modeling of the \textit{Ln}N with a $\sim$3\% concentration of \textit{Ln} vacancies, including computation of the V$_{\text{cat}}$ breathing-mode frequency, for comparison with the Raman data. Section IV summarises the entire study. 

\section{Raman spectroscopy}

The preparation of the \textit{Ln}N powders and thin films is discussed in detail in earlier papers~\cite{VanKoughnet2023,Kneisel2024}. The films were grown in a conventional MBE system, exploiting the facility of the clean lanthanide surface to break the N$_{\text{2}}$ triple bond~\cite{Ullstad2019,chan2020facile,chan2023}. Polycrystalline \textit{Ln}N films were grown on c-plane sapphire to a thickness of $\sim$100~nm and capped with passivating AlN. As mentioned before, powders were prepared by ball milling in a nitrogen-loaded glove box~\cite{Kneisel2024}.

A Jobin-Yvon LabRam was employed for Raman studies, using a variety of excitation lasers~\cite{VanKoughnet2023,Kneisel2024}. Film samples were simply placed on the microscope stage, but powders required to be introduced into a sealed passivating chamber fitted with a transparent window before removing them from the nitrogen glove box~\cite{Kneisel2024}. That chamber was then set on the microscope stage as for the films, and Raman measurements performed through the window. 

\begin{figure}
    \centering
    \includegraphics[width=\linewidth]{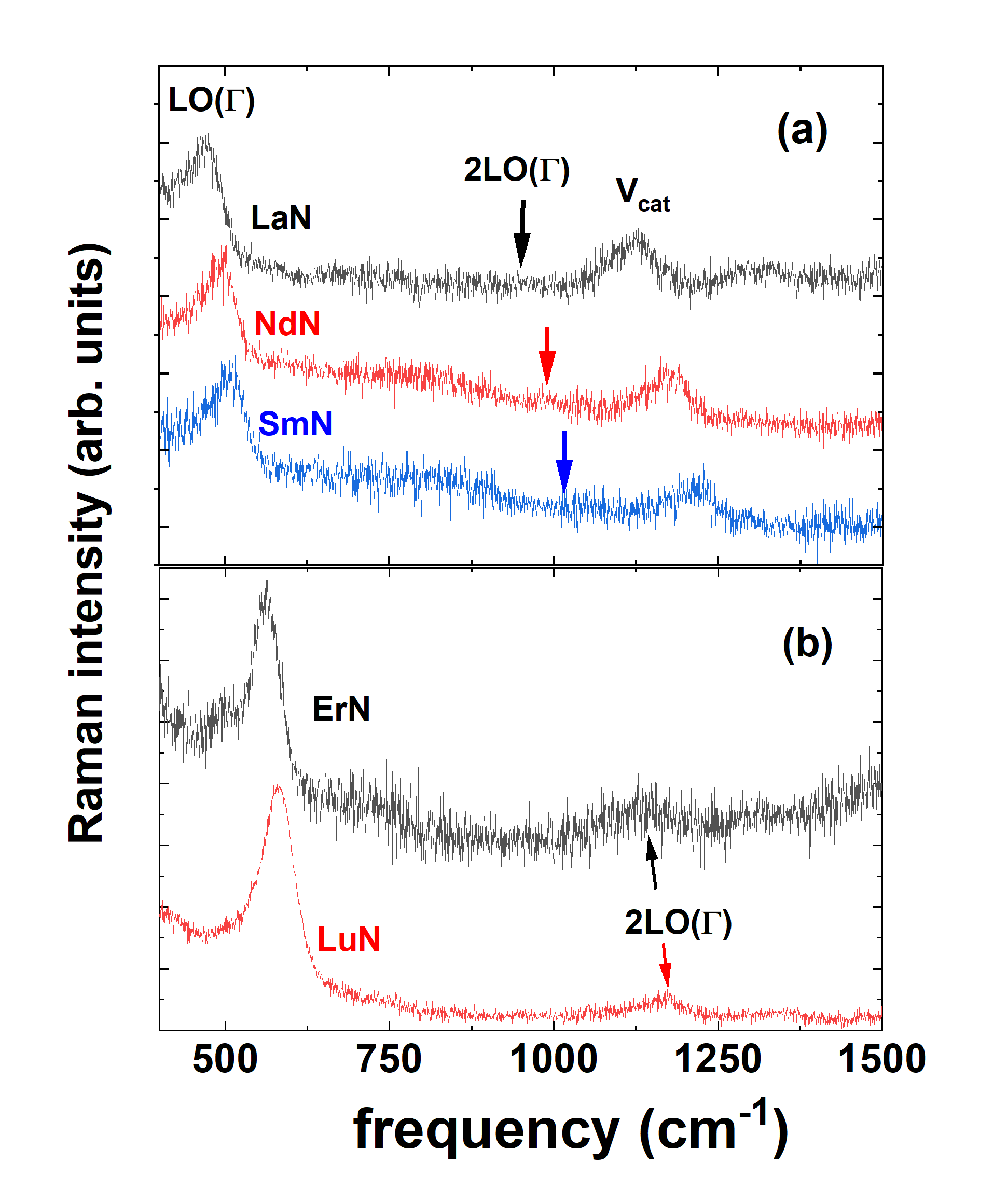}
    \caption{Raman spectroscopy of \textit{Ln}N powders, with (a) light \textit{Ln}N species exhibiting V$_{\text{cat}}$ peaks, and (b) heavy \textit{Ln}N species exhibiting 2LO($\Gamma$) peaks. The strongest line is labeled as LO($\Gamma$), and if its overtone appeared it would be at the 2LO($\Gamma$) labeled position. Note that the feature labeled as V$_{\text{cat}}$ is $\sim$200~cm$^{-1}$ above 2LO($\Gamma$).}
  \label{fig:powders}
\end{figure}

Raman spectroscopy was undertaken on powders of 8 among the 15 \textit{Ln}N: LaN, NdN, SmN, GdN, TbN, DyN, ErN, and LuN~\cite{Kneisel2024}. All showed the \textit{Ln}N LO($\Gamma$) Raman signal. The V$_{\text{cat}}$ line was present in the three lightest (LaN, NdN, SmN) as seen in Fig.~\ref{fig:powders}a. The data show a clear decrease of the V$_{\text{cat}}$ and increase of the 2LO($\Gamma$) signal strength with rising \textit{Ln} mass. Only for the two heaviest (ErN, LuN) did we find the 2LO($\Gamma$) overtone (Fig.~\ref{fig:powders}b), but note that the signal to noise ratio obscures the weak overtone signal in the rest of the powders. 

\begin{figure}
    \centering
    \includegraphics[width=\linewidth]{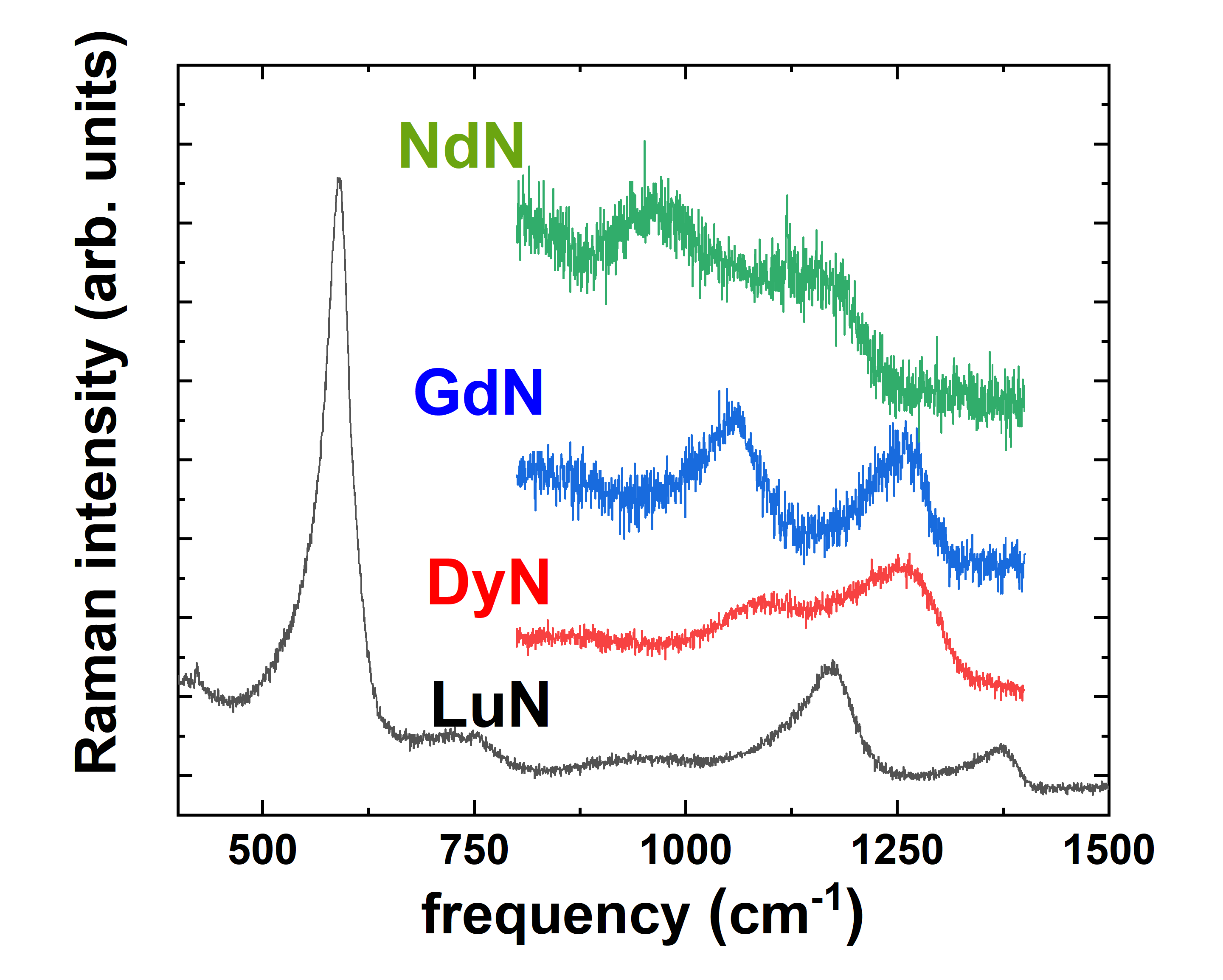}
    \caption{Raman spectra from LuN, DyN, and GdN films grown under activated nitrogen from an ion source and a NdN film grown in N$_{\text{2}}$. The pair of lines, 2LO($\Gamma$) and $\sim$200~cm$^{-1}$ higher V$_{\text{cat}}$, are seen clearly in all four films.}
    \label{fig:Vcat lines}
\end{figure}

Turning to thin films, the V$_{\text{cat}}$ feature is absent on GdN and LuN films grown as usual in the presence of pure nitrogen, N$_{\text{2}}$, but GdN, DyN and LuN films grown in the presence of $\sim$120~eV nitrogen ions show exactly that same feature $\sim$200~cm$^{-1}$ higher (Fig.~\ref{fig:Vcat lines}). Evidently, ions were required for the formation of cation vacancies responsible for the breathing-mode vibration in these films. However, the feature appears also in the spectrum of one in the series, NdN, grown in an atmosphere of 4$\times$10$^{-6}$ mbar of unactivated N$_{\text{2}}$ gas, showing both 2LO($\Gamma$) and V$_{\text{cat}}$. 

\begin{figure}
    \centering
    \includegraphics[width=\linewidth]{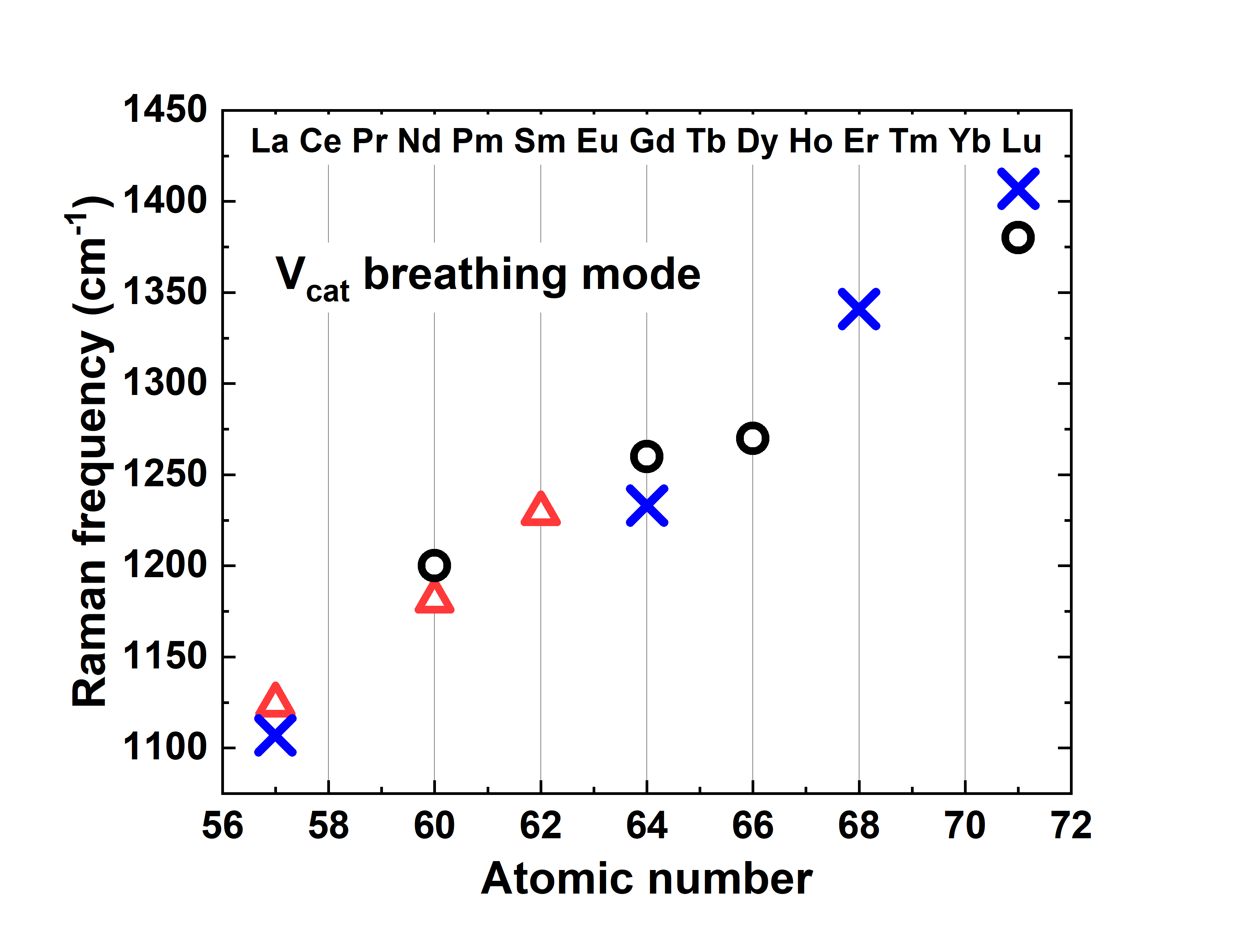}
    \caption{Series of V$_{\text{cat}}$ Raman frequencies with Z. Data from powders are shown as open red triangles, from thin films as open black circles and the blue crosses are as computed.}
    \label{fig:freq_vs_Z}
\end{figure}

Assembling the V$_{\text{cat}}$ frequencies across the full set of powder and film data, we show in Fig.~\ref{fig:freq_vs_Z} their run with atomic number. There is a steady Raman-frequency rise with increasing Z, as might be anticipated based on the reduced lattice constant across the \textit{Ln}N series.

\section{DFT modeling}

The DFT calculations were conducted using \texttt{Quantum} \texttt{ESPRESSO}~7.3 with pseudopotentials developed by Topsakal and Wentzcovitch, and Dal Corso~\cite{giannozzi2009quantum,giannozzi2017advanced,giannozzi2020quantum,topsakal2014accurate,dal2014pseudopotentials}. The calculations were performed at the generalized gradient approximation-level using the Perdew-Burke-Ernzerhof (PBE) exchange-correlation potential approximation~\cite{larson2007electronic}. They were conducted with full spin polarization on the two-atom primitive unit cells of the Fm$\overline{\text{3}}$m group. The primitive unit cells were converged using a wavefunction energy cutoff of 60~Ry (816~eV) and a 16$\times$16$\times$16 k-point mesh, and the total pressure relaxed below 0.1~Pa.

Rotationally-invariant Hubbard~$U$ parameters with double-counting accounted for in the fully-localized limit were applied to the \textit{Ln}~4$f$ and N~2$p$ states in order to correct the description of the electronic structure~\cite{dudarev1998electron}. The Hubbard parameters (i) separate the \textit{Ln}~4$f$ levels from the Fermi energy and (ii) open the band gaps, similar to the application of the Hubbard~$U$ applied to the \textit{Ln}~5$d$ states, further described elsewhere~\cite{holmes2021electronic}. The choice of Hubbard parameters was made with recourse to the experimental band structure probes where available, and otherwise scaled relative to the $U_{f}$ parameter fitted to the experimental Gd $4f$ state depth of $7.8$~eV below the valence band edge in GdN~\cite{leuenberger2005gdn,trodahl2007ferromagnetic,devese2022probing}. The resulting combination of Hubbard parameters are summarized in Table~\ref{tab:computation}.

\begin{table}
\caption{\label{tab:computation}Summary of Hubbard parameters used for the DFT+$U_{f}^{\text{\textit{Ln}}}$+$U_{p}^{\text{N}}$ calculations and the derived structural, thermodynamic, and vibrational properties of LaN, GdN, ErN, and LuN.}
\begin{ruledtabular}
\begin{tabular}{c c c c c}
 Property & LaN & GdN & ErN & LuN \\
 \hline
 $U_{f}^{\text{\textit{Ln}}}$ (eV) & 1.6 & 10.25 & 8.34 & 7.87 \\
 $U_{p}^{\text{N}}$ (eV) & 4.97 & 5.12 & 2.32 & 4.41 \\
 $E_{\text{coh.}}$ (eV) & 7.80 & 8.44 & 8.92 & 9.00 \\ 
 $E$(V$_{\text{cat}}$) (eV) & 7.76 & 9.47 & 10.63 & 10.99 \\ 
 a (nm) & 0.5320 & 0.4996 & 0.4845 & 0.4772 \\
 LO($\Gamma$) (cm$^{-1}$) & 482 & 540 & 547 & 587 \\
 2LO($\Gamma$) (cm$^{-1}$) & 964 & 1080 & 1094 & 1174 \\
 Breath. (cm$^{-1}$) & 1107 & 1233 & 1341 & 1401 \\
 \hline
\end{tabular}
\end{ruledtabular}
\end{table}

To understand the relationship between the concentration of V$_{\text{cat}}$ for the different \textit{Ln}N compounds, defect formation energy calculations were conducted using the $3\times3\times3$ supercell of the primitive unit cell as a starting structure~\cite{porat2024tuneable}. The same numerical cutoffs were used for these computations with a 3$\times$3$\times$3 k-point mesh. The total energies of the defected supercell, primitive unit cell, and isolated reference atoms were used in conjunction with the experimental elemental binding energies to determine the \textit{Ln}N cohesive energies ($E_{\text{coh.}}$), summarized in Table~\ref{tab:computation}~\cite{kittel2018introduction}. The defect formation energies for V$_{\text{cat}}$ were computed after conducting pressure and force-relaxation after removing one cation from the $54$-atom supercell, corresponding to a cation vacancy concentration of $\sim$3\%. The pressure on the supercell and the forces on each atom were lowered to within 0.5~Pa and 10~meV~{\AA}$^{-1}$, respectively. While we do not investigate charged supercells, the resulting neutral V$_{\text{cat}}$ energies are similar to those evaluated using the Heyd-Scuderia-Ernzerhof (HSE06) exchange-correlation functional for ScN and LaN~\cite{kumagai2018point,deng2021semiconducting}. We note that due to the large $E$(V$_{\text{cat}}$) for all \textit{Ln}N we expect the concentration of cation vacancies in experiment to be much less than the modeled $3$\%.

Density functional perturbation theory (DFPT) computations were conducted on DFT+$U_{p}$+$U_{f}$-corrected \textit{Ln}N primitive unit cells to investigate their vibrational structures~\cite{floris2011vibrational,floris2020hubbard}. We note that for LaN the phonon dispersion exhibits imaginary phonon frequencies when using the Fm$\overline{\text{3}}$m group, remedied by use of the P1 group instead, both shown in SM Fig. S3~\cite{markwitz2024supplemental}. We randomly included small distortions the lattice vectors and angles and allowed the crystal structure to relax both to minimize the total energy of the LaN primitive unit cell. These computations found similar results to Chen \textit{et al.}~\cite{chen2021lan}, employing just the PBE exchange-correlation potential approximation without a Hubbard parameter correction. A table of quantities which compares the derived P1 lattice vectors, angles, and atomic positions to those of the Fm$\overline{\text{3}}$m group for LaN are in the supplementary information SM Table~SI~\cite{markwitz2024supplemental}. The derived quantities (lattice constant and LO($\Gamma$) phonon frequency) based on the choice of symmetry group for LaN varied less than 4\%. LO-TO splitting was considered for these computations via the acoustic sum rule, requiring the computation of the high-frequency permittivity and the Born effective charges. These computations were conducted on defect-free primitive unit cells of each \textit{Ln}N at their relaxed lattice constants, the resultant phonon dispersions and density of modes for each \textit{Ln}N can be found in SM Fig.~S2~\cite{markwitz2024supplemental}. The high-frequency permittivities (Born effective charges) were computed to be $7.464$ ($\pm4.03q$), $8.342$ ($\pm4.27q$), $9.668$ ($\pm4.26q$), and $7.311$ ($\pm3.93q$), for LaN, GdN, ErN, and LuN, respectively, where $q$ is the elementary charge. The resulting equilibrium lattice constants and the corresponding LO($\Gamma$) phonon mode frequencies are summarized in Table~\ref{tab:computation}, all of which lay close to the corresponding experimental values~\cite{Natali2013,anton2023growth,devese2022probing,Kneisel2024,VanKoughnet2023,Granville2009}. The calculated lattice constants using the DFT+$U_{p}$+$U_{f}$ approximation are within 0.5\% of the experimental values as seen in SM Table~SII~\cite{markwitz2024supplemental}, and the LO($\Gamma$) phonon modes differ by no more than 5.5\%, the agreement between computation and experiment improving with increasing atomic number. 

We now move on to relaxed V$_{\text{cat}}$-defected 53-atom 3$\times$3$\times$3 supercells. The first-nearest-neighbor N atoms are displaced away from the vacancy site as a result of the relaxation by 29.3, 18.6, 17.9, and 15.7~pm, for LaN, GdN, ErN and LuN, respectively, corresponding to an increase of the V$_{\text{cat}}$-N spacing by 11.0\%, 7.0\%, 7.3\%, and 6.6\%. This is due to the attraction of those neighbors to non-vacant \textit{Ln} opposing the vacancies. An increase of the lattice parameter local to the vacancy was noted also when introducing V\textsubscript{an} to SmN~\cite{holmes2021electronic} and GdN~\cite{holmes2024spin}. The computed defect formation energy of V$_{\text{cat}}$ in metal-poor deposition conditions is included in Table~\ref{tab:computation}. This suggests that lighter \textit{Ln}N species can be expected to have a higher concentration of V$_{\text{cat}}$, as defects with greater formation energies occur at a lower concentration, as observed in the powder data where only the light \textit{Ln}N show the V$_{\text{cat}}$ Raman signal.

\begin{figure}
    \centering
    \includegraphics[width=\linewidth]{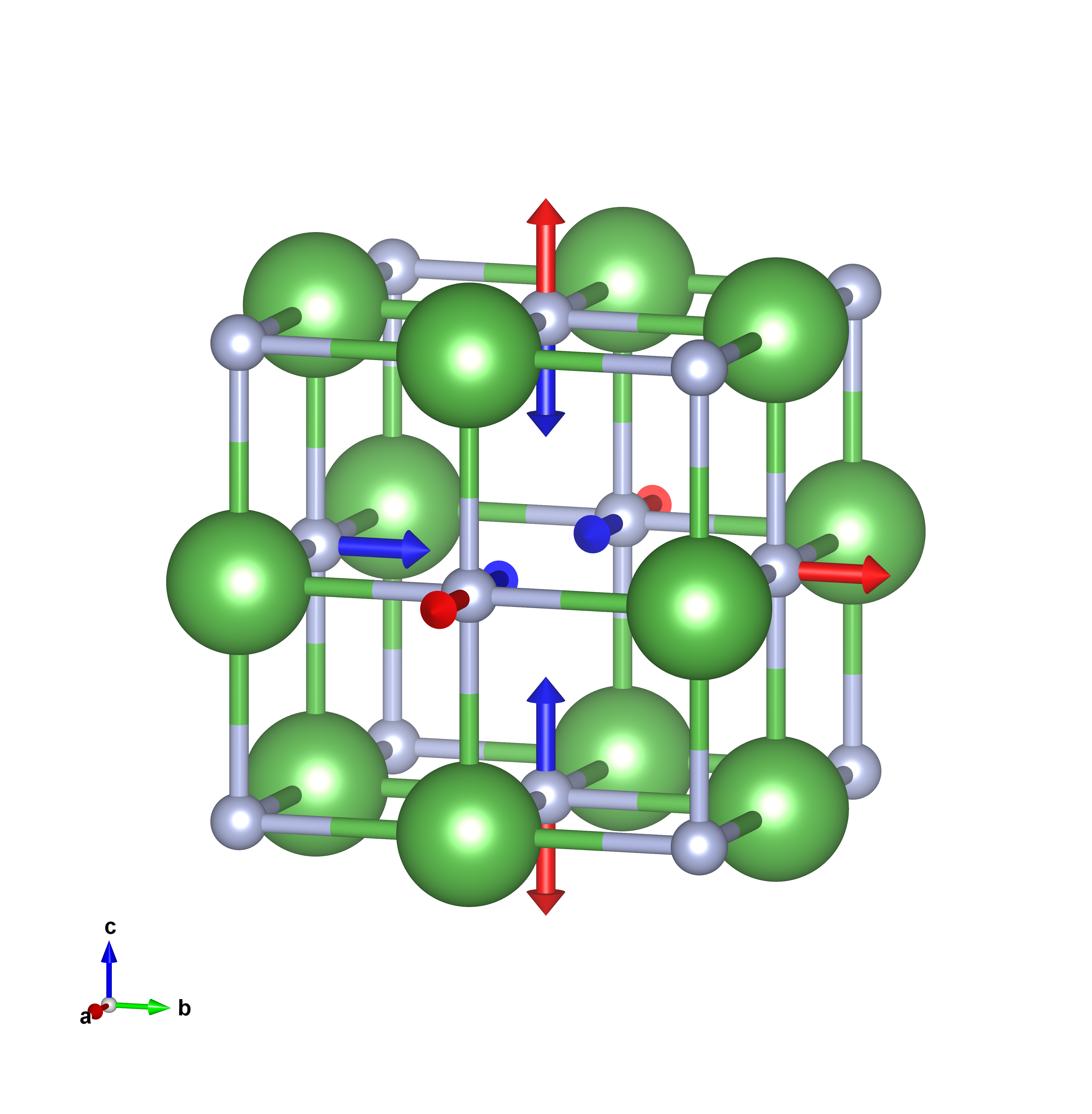}
    \caption{Conventional \textit{Ln}N rock-salt unit cell containing one V$_{\text{cat}}$ (green, \textit{Ln} and grey, N). The displacement vectors designate the vibrational motion of the breathing mode (red is outward, and blue is inward).}
\label{fig:displacementDiagram}
\end{figure}

The breathing mode frequencies were found by calculating the variation in the total energy of the relaxed V$_{\text{cat}}$-defected supercell upon the positional perturbation of the six nitrogen atoms neighboring the vacancy. The atomic positions were not relaxed for this series of calculations. See Fig.~\ref{fig:displacementDiagram} for a visual representation of the distortion of the six neighboring nitrogen atoms to the vacancy site made using VESTA~\cite{momma2008vesta}, while the rest of the atomic positions are clamped. A third-order polynomial was fitted to the displacement-total energy data and the prefactor to the second order component used to extract the breathing mode frequency, the fit quality included in SM Fig.~S3~\cite{markwitz2024supplemental}. More details on the method for this calculation can be found in the supplemental information. The breathing mode frequencies are summarized in Table~\ref{tab:computation}, finding that those modes lay $\sim200$~cm$^{-1}$ above the calculated 2LO($\Gamma$) overtone frequency of its host \textit{Ln}N. They are also compared to experimental data in Fig.~\ref{fig:freq_vs_Z}, finding excellent agreement for LaN, GdN, and LuN, and suggest the breathing mode frequency for ErN, data for which has not yet been realized experimentally. Note, however, that this breathing mode models the displacement of only the first-nearest-neighbor N ions, ignoring even the second-nearest-neighbor \textit{Ln} ions. A fuller theoretical analysis and computational model which does not demand fixed-site second-nearest-neighbors (and so on) is expected to soften the calculated breathing mode frequency by no more than 2.5\%. This estimate originates from the mass difference of the shell of six nearest-neighbor N atoms and the second shell of 12 second-nearest-neighbor \textit{Ln} atoms. Finally, the presence of these breathing modes as measured by Raman spectroscopy directly implies the presence, and by extension, the concentration of cation vacancies in the rock salt-structured \textit{Ln}N.

\section{Summary}

We have identified a Raman signature of cation vacancies in the \textit{Ln}N and explored the preparation of such defects across the width of the 15-member series. The assignment of the feature as arising from \textit{Ln} vacancies is well-supported by a DFT computation in the usual rock-salt \textit{Ln}N structure decorated with $\sim$3\% cation vacancies that reproduces the atomic-number dependence of the breathing-mode frequency. The vacancies are found in (i) MBE films exposed to $\sim$120~eV ionic nitrogen during growth and (ii) powders formed by ball milling pure metals of the light \textit{Ln} (La-Sm) in a nitrogen atmosphere. Interestingly, the breathing mode also appears in NdN films grown with molecular nitrogen, suggesting that NdN may have a particular propensity toward forming cation vacancies. The discovery of the capacity to incorporate \textit{Ln} vacancies into \textit{Ln}N opens the door to intrinsic acceptor doping to compensate the \textit{anion} vacancies that are a common feature among the \textit{Ln}N series of intrinsic ferromagnetic semiconductors. 

\section{Acknowledgments}

The authors have benefitted from many informative and encouraging discussions with Bob Buckley, Jackson Miller and Simon Granville, and technical/operational support from Tane Butler. This research was supported by Quantum Technologies Aotearoa (contract UOO2347), a research programme of Te Whai Ao — the Dodd Walls Centre, and by the New Zealand Endeavour fund (contract RSCHTRUSTVIC2447). Travel support was received from a QuantEmX grant GBMF9616 from ICAM and the Gordon and Betty Moore Foundation. M. Markwitz thanks the MacDiarmid Institute for funding his role on this project. The MacDiarmid Institute is supported under the New Zealand Centres of Research Excellence programme. The computations were performed on R$\overline{\text{a}}$poi, the high performance computing facility of Victoria University of Wellington.

\bibliography{Ram}

\end{document}